\documentclass[
 aps,
 10pt,
 final,
 notitlepage,
 oneside,
 twocolumn,
 nobibnotes,
 nofootinbib,
 superscriptaddress,
 noshowpacs,
 centertags]
{revtex4}

\usepackage[dvips]{graphicx}
\usepackage{url}
\usepackage{amsmath}

\def\squareforqed{\hbox{\rlap{$\sqcap$}$\sqcup$}}

\def\sq{\ifmmode\squareforqed\else{\unskip\nobreak\hfil
\penalty50\hskip1em\null\nobreak\hfil\squareforqed
\parfillskip=0pt\finalhyphendemerits=0\endgraf}\fi}

\def\utw{\smash{\rlap{\lower5pt\hbox{$\sim$}}}}

\def\udtw{\smash{\rlap{\lower6pt\hbox{$\approx$}}}}

\def\diameter{{\ifmmode\mathchoice
{\ooalign{\hfil\hbox{$\displaystyle/$}\hfil\crcr
{\hbox{$\displaystyle\mathchar"20D$}}}}
{\ooalign{\hfil\hbox{$\textstyle/$}\hfil\crcr
{\hbox{$\textstyle\mathchar"20D$}}}}
{\ooalign{\hfil\hbox{$\scriptstyle/$}\hfil\crcr
{\hbox{$\scriptstyle\mathchar"20D$}}}}
{\ooalign{\hfil\hbox{$\scriptscriptstyle/$}\hfil\crcr
{\hbox{$\scriptscriptstyle\mathchar"20D$}}}}
\else{\ooalign{\hfil/\hfil\crcr\mathhexbox20D}}%
\fi}}

\newcommand{\ab}{Astrophysical Bulletin }

\newcommand{\aaa}{Astron. and Astrophys. }
\newcommand{\aap}{Astron. and Astrophys. }

\newcommand{\aaps}{Astron. and Astrophys. Suppl. }

\newcommand{\aj}{Astron.~J. }

\newcommand{\mnras}{Monthly Notices Royal Astron. Soc. }

\newcommand{\pasp}{Publ. Astron. Soc. Pacific }

\begin{document}


%


 \title{A comprehensive study of 94 open clusters based on the data from IPHAS,
GAIA DR2, and other sky surveys }

\author{\firstname{L.~N.}~\surname{Yalyalieva}}
\email{yalyalieva@yandex.ru}
\affiliation{Sternberg Astronomical Institute, M.~V.~Lomonosov Moscow State University,
Universitetskii pr. 13, Moscow, 119992 Russia} \affiliation{Lomonosov Moscow State University, Faculty of
Physics, 1, bld.2, Leninskie Gory, Moscow,  119992, Russia}

\author{\firstname{A.~A.}~\surname{Chemel}}
\affiliation{Lomonosov Moscow State University, Faculty of
Physics, 1, bld.2, Leninskie Gory, Moscow,  119992, Russia}

\author{\firstname{E.~V.}~\surname{Glushkova}}
\email{elena.glushkova@gmail.com}
\affiliation{Sternberg Astronomical Institute, M.~V.~Lomonosov Moscow State University,
Universitetskii pr. 13, Moscow, 119992 Russia} \affiliation{Lomonosov Moscow State University, Faculty of
Physics, 1, bld.2, Leninskie Gory, Moscow,  119992, Russia}

\author{\firstname{A.~K.}~\surname{Dambis}}
\email{dambis@yandex.ru}
\affiliation{Sternberg Astronomical Institute, M.~V.~Lomonosov Moscow State University,
Universitetskii pr. 13, Moscow, 119992 Russia}

\author{\firstname{A.~D.}~\surname{Klinichev}}
\affiliation{Lomonosov Moscow State University, Faculty of
Physics, 1, bld.2, Leninskie Gory, Moscow,  119992, Russia}

\begin{abstract}
We determine the color excesses, photometric distances, ages,  astrometric parallaxes and proper motions
for 94 open clusters in the northern part of the Milky Way. We estimate the color excesses and photometric
distances based on the data from IPHAS photometric survey of the northern Galactic plane using individual
total-to-selective extinction ratios  $R_r$~=$A_r$/$E_{r-i}$ for each cluster computed via the color-difference
method based on IPHAS $r$, $i$, and $H_{\alpha}$-band, 2MASS $J, H$, and $K_s$-band,
WISE $W1$-band, and Pan-STARRS $i, z$, and $y$-band data. The inferred $R_r$  values
vary significantly from cluster to cluster spanning the $R_r$~=3.1--5.2 interval with a mean and standard deviation
equal to $<R_r>$~=3.99 and $\sigma R_r$~=0.34, respectively. 
We identified cluster members using (1) absolute proper motions determined from individual-epoch positions of stars retrieved 
from IPHAS, 2MASS, URAT1, ALLWISE, UCAC5, and Gaia DR1 catalogs and positions of stars on individual Palomar Sky Survey 
plates reconstructed based  on the data provided in  USNO-B1.0 catalog and (2) absolute proper motions
provided in Gaia DR2 catalog, and computed the average Gaia DR2 trigonometric parallaxes and proper motions of the
clusters. The mean formal error of the inferred astrometric parallaxes of clusters is of about 
7~microarcseconds, however, a comparison of astrometric and photometric parallaxes of our cluster sample
implies that Gaia DR2 parallaxes are, on the average, systematically underestimated by  45~$\pm$~9~microarcseconds.
This result agrees with estimates obtained by other authors using other objects. At the same time, we find our
photometric distance scale to be correct within the quoted errors (the inferred correction factor is equal to unity
to within a standard error of 0.025).

\end{abstract}

\maketitle

 \section{INTRODUCTION}
\label{intro} 
The number of sky surveys providing various data about the properties of a huge number of objects
has increased substantially in recent decades. When using photometric catalogs most of the
researchers employ them only as sources of star magnitudes in various passbands. However, all
such catalogs also contain equatorial coordinates of the objects, which can be used for
computing both relative and absolute proper motions of stars over a long time interval, which
is especially important for binary systems. A recent study by Chemel et al.~\citep{Chemel} represents
a successful case of the use of six major sky surveys for determining absolute proper motions of
115 Galactic globular clusters.

In this study we attempt to determine the proper motions of stars using position data from seven photometric
sky surveys in the neighborhood of Galactic open clusters observed within the framework of 
IPHAS survey~\citep{IPHAS1,IPHAS2}. We do it to identify cluster members and apply to them 
the algorithm for estimating the basic open cluster parameters via a modified Q-method as 
described by Dambis et al.~\citep{Dambis}.
Although young open clusters are the most suitable Galactic disk tracers, their basic parameters
(distances, ages, and color excesses) still needs to be refined. One of the two most popular
sources of information about open cluster parameters - the catalog of Dias et al.~\citep{Dias} 
is a compilation and contains the data of various accuracy and reliability for about 2000 objects. 
Although the authors of another data source - Kharchenko et al.~\citep{Kharchenko} - determined
the parameters of more than 3000 clusters by applying the same technique to  2MASS photometry~\citep{TMASS}, 
their estimates are reliable only for nearby and old clusters whose Hertsprung--Russel diagrams 
show a conspicuous clump at the base of the red giant branch. For all other
open clusters the parameter estimates based on relatively shallow  $J$,$H$,$K{_S}$ photometry
is unreliable and this especially concerns distances, which are determined by fitting an isochrone
to the Hertsprung--Russel diagram, because for the photometric bands considered 
the main sequence is practically vertical for young clusters located beyond 1~kpc.

In the context of this study of special  interest is investigation
of the extinction law toward each cluster and a comparison of the average cluster parallaxes
inferred from GAIA DR2 data~\citep{Gaia, DR2} with the corresponding photometric parallaxes
determined from IPHAS photometry~\citep{IPHAS1, IPHAS2}.

\section{EXTRACTION OF CLUSTER DATA FROM CATALOGS}
\label{Data}

\subsection{Color excess, distance, and age}
\label{parameters}

We estimate the physical parameters of clusters (color excesses, heliocentric distances, and ages) from
Sloan  $r$- and $i$-band and narrow-band $H_{\alpha}$ photometry from IPHAS catalog~\citep{IPHAS1, IPHAS2}. 
This survey, which was carried out on the 2.5m Isaac Newton Telescope
(INT) in the northern part of the sky,  covers the Galactic longitudes $l$=~30$^o$ - 215$^o$
and latitudes $|b|$<~5$^o$ down to a limiting magnitude of 21.2 mag,
20.0 mag, and 20.3 mag in the $r$, $i$, and $H_{\alpha}$ filters, respectively.
Dambis et al.~\citep{Dambis} modified the classical $Q$-method of Johnson and Morgan~\citep{Johnson}
and applied it to  $r$-, $i$-, and $H_{\alpha}$-band magnitudes to construct the reddening-free
index  $H_{\alpha} index$ =~($0.755r$ + $0.245i$
- $H_{\alpha}$). Like in the paper of Dambis et al.~\citep{Dambis}, here we use the
($H_{\alpha} index$, $r$-$i$) diagram to determine the color excess $E(r-i)$. 
The reddening lines in this diagram are parallel to the horizontal axis because
the combination ($0.755r$ + $0.245i$) linearly interpolates between the corresponding
photometric passbands and sort of imitates a broadband  $H_{\alpha}$ filter so that
$H_{\alpha} index$ does not depend on reddening. 
In the ($r$-$i$, $H_{\alpha} index$) diagram Padova PARSEC isochrones~\citep{Chen1, Chen2, Tang} with
ages in the  $log(t)$=~$6.0-8.5$ interval all
have a minimum at the same $(r-i)$ value and its position is practically
independent of metallicity. This is because the strength of $H_{\alpha}$ absorption line
is the highest in  $A0-A2$-type stars.
Therefore this method is applicable only to relatively young clusters ($log(t)<8.5$)
whose main sequences still contain stars of such spectral types. We thus determined the color excess $E(r-i)$ 
by shifting the isochrone (at the start we set the age equal to $log(t)$=~7.2) along the horizontal
axis. We then determined the cluster distance modulus from the  ($r-i$,~$r$),
($H_{\alpha}index$,~$r$), and ($H_{\alpha}index$,~$r$-~$3.98(r-i)$) diagrams and estimated the cluster
age only from the  ($r-i$,~$r$) and ($H_{\alpha}index$,~$r$-~$3.98(r-i)$) diagrams. We adopted the $r$- and $i$-band
magnitudes for stars brighter than $r$=~13, $i$=~12 from  APASS survey~\citep{APASS1, APASS2} and transformed
them to the IPHAS photometric system by formulas from~\citep{IPHAS1} because images of these stars are saturated 
in the IPHAS frames. We described the entire multi-stage procedure in our previous paper~\citep{Dambis}.

\begin{figure}
\includegraphics[width=1.0\linewidth]{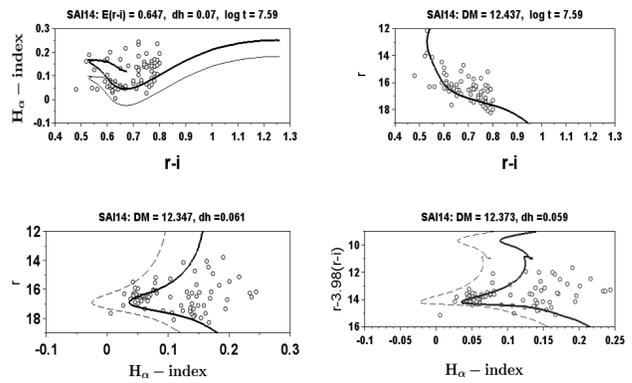}
 \caption{The ($r$-$i$, $H_{\alpha} index$) diagram (top left) and the color-magnitude
diagrams ($r$-$i$, $r$) (top right), ($H_{\alpha} index$, $r$) (bottom left),
and ($H_{\alpha} index$, $r$-~$3.98(r-i)$) (bottom right) for all stars in the field of  SAI~14 
cluster within 5~arcmin from its center. The thick solid line shows the  $log(t)=7.59$
isochrone shifted in accordance with the inferred color excess and distance modulus. 
The thin solid or dashed line shows the isochrone before the final shift taking into account
the zero-point difference between the theoretical isochrone and observed main sequence of
the cluster.}
 \label{fig:param_all}
\end{figure}

The top left panel in Fig.~\ref{fig:param_all} shows the two-color diagram with the
$log(t)$=~7.59 isochrone (the thin line) shifted along the horizontal axis by 0.647 
to fit the main sequence of the $SAI~14$ cluster. The thick line shows the same isochrone additionally
shifted along the vertical axis to take into account a small systematic zero-point difference between
the theoretical isochrones and actually observed sequences. The other three panels in Fig.~\ref{fig:param_all} 
show isochrones shifted by the distance modulus (the shift in the two lower panels
is shown by the dashed lines). The final shift in the two lower panels takes into account the zero-point difference
mentioned above.

We estimated all parameters in automatic way using the maximum-likelihood based algorithm proposed by 
Naylor et al.~\citep{Naylor, Jeffries, Mayne} and averaged the distance moduli inferred from three 
different diagrams.

We determined the distances, color excesses, and ages for 108 open clusters from the catalog of
Dias et al.~\citep{Dias}, for which the data were available in the IPHAS survey~\citep{IPHAS1, IPHAS2} 
and whose  ($H_{\alpha} index$, $r$-$i$) showed a minimum.

However, to more confidently estimate the parameters we need to separate cluster members from field stars. 
To this end we determined the proper motions of stars in the fields of 108 clusters.

\subsection{Proper motions and cluster membership of stars}
\label{motions}

To derive proper motions of stars we used the results of major photographic surveys 
POSS-I, POSS-II, SERC-J, SERC-EJ, ESO-R,
AAO-R, SERC-ER, and SERC-I spanning the period from 1949 through 2002 and incorporated 
into  USNO-B1.0~\citep{USNOB1} catalog, and the large-scale sky surveys UCAC5~\citep{UCAC5} 
(1998-2004), 2MASS~\citep{TMASS} (1997--2001), WISE~\citep{WISE1, WISE2}
(2010-2011 with a mean epoch of 2010.5589), URAT1~\citep{URAT} (2012--2015), 
IPHAS~\citep{IPHAS1, IPHAS2} (2003--2012), and Gaia
DR1~\citep{Gaia} (with a mean epoch of 2015.0), which contain sufficiently precise positional data. 
We first  had to cross identify stars over  the entire set of catalogs.

We had at our disposal such tools for operating with large astronomical datasets as  Aladin
interactive sky atlas~\citep{ALADIN}, as well as STILTS~\citep{STILTS},
TOPCAT~\citep{TOPCAT}, and С3~\citep{C3} packages. They can be used to perform pairwise or even
multiple cross identification of catalogs, but are not very convenient for regular work because
the user has each time to repeat the same sequence of operations (define the set of catalogs 
employed, cross-match conditions, formats of output files, etc). To simplify this procedure, we 
developed Crossmatch program,  which perform cross-matching of stars over arbitrary set of
catalogs in a field of arbitrary radius around the given center, is easy to use and flexible. 
Crossmatch program is written in java and is available as a jar-file, which can be run from the
command line. A substantial part of operations with tables described in Crossmatch algorithm is
performed using the capabilities of  STILTS program~\citep{STILTS}. The  source code of the
program is available as a single archive file \verb|Crossmatch_4.3.0.zip|
at:
\newline
\verb|www.sai.msu.ru/groups/cluster/cl/crossmatch/|

We used this program to mutually cross identify the catalogs considered within 30~arcmin from each open
cluster with an identification radius of 1 or 2~arcsec. As a result, we obtained up to 13 positions
per star. A detailed description of the procedure that we used to compute proper motions
can be found in the paper by Chemel et al.~\citep{Chemel}, which is dedicated to the study
of the kinematics of Galactic globular clusters. All proper motions are in the system
defined by UCAC5 catalog and the accuracy of individual proper motions is of about 1--2~mas/yr.

We then used the inferred proper motions to identify members of each cluster via the
commonly used Sanders's method~\citep{Sanders}.
This method allows determining the mean proper motion of cluster and field stars, 
$\left( \mu_{\alpha}^{*}, \mu_{\delta}^{*} \right)$ and $\left( \mu_{\alpha}, \mu_{\delta} \right)$, 
the dispersions of the proper motions of cluster stars --  $\left( \sigma_{\alpha}^{*}, 
\sigma_{\delta}^{*} \right)$ -- and the dispersion of proper motions of field stars,
$\left( \sigma_{\alpha},
\sigma_{\delta} \right)$, as well as the fraction of cluster members $\left( N \right)$. 
We implemented this method assuming that the velocity distribution of cluster stars is isotropic,
$\left(\sigma_{\alpha}^{*}=\sigma_{\delta}^{*}=\sigma^{*} \right)$, and turned the coordinate frame 
so as to align the axes of the ellipse of the distribution of proper motions of field stars 
with the coordinate axes.

Probability density functions of cluster members ($F^{*}$) and field stars ($F^{f}$) have the following form:

\begin{center}
\begin{align}
            F_i^*=\frac{N}{2\pi\sqrt{{\sigma^{*}}^2+{\varepsilon_\alpha^{i}}^2}\sqrt{{\sigma^{*}}^2+{\varepsilon_\delta^{i}}^2}}\cdot\nonumber\\
            \cdot\exp\left( -0.5\left[ \frac{{\left( \mu_\alpha^i-\mu_\alpha^* \right)}^2}{{\sigma^{*}}^2+{\varepsilon_\alpha^{i}}^2}+\frac{{\left( \mu_\delta^i-\mu_\delta^* \right)}^2}{{\sigma^{*}}^2+{\varepsilon_\delta^{i}}^2} \right]
            \right)
  \end{align}
  \begin{align}
            F_i^f=\frac{1-N}{2\pi\sqrt{{\sigma_\alpha}^2+{\varepsilon_\alpha^{i}}^2}\sqrt{{\sigma_\delta}^2+{\varepsilon_\delta^{i}}^2}}\cdot\nonumber\\
            \cdot\exp\left( -0.5\left[ \frac{{\left( \mu_\alpha^i-\mu_\alpha \right)}^2}{{\sigma_\alpha}^2+{\varepsilon_\alpha^{i}}^2}+\frac{{\left( \mu_\delta^i-\mu_\delta \right)}^2}{{\sigma_\delta}^2+{\varepsilon_\delta^{i}}^2} \right]
            \right),
        \end{align}
     \end{center}

where $\mu_\alpha^i$ and $\mu_\delta^i$, $\varepsilon_\alpha^i$, and
$\varepsilon_\delta^i$ are  the proper-motion components and their standard errors for $i$th
star of the sample. The full probability density function then has the form $F_i=F_i^*+F_i^f$.

The proper motions of individual stars are independent of each other and therefore we
compute the likelihood function as the product of probability density  functions 
over all stars of the sample: $Q=\prod_i{F_i}$.
In accordance with the maximum likelihood principle we consider the true parameter values to be those 
corresponding to the most likely distribution, i.e., at the maximum of the likelihood function. 
In practice, however, it is more convenient to search for the minimum of a function of the
following form:

\begin{center}
        $\displaystyle L=-\log{Q}=-\sum\limits_i{\log{F_i}}$
    \end{center}.

We minimized this function via the conjugate gradient method and estimated the cluster membership probability
of each star by the formula:
\begin{center}
        $\displaystyle P_i=\frac{F_i^*}{F_i}$
    \end{center}.

\begin{figure}[]
\includegraphics[width=\linewidth]{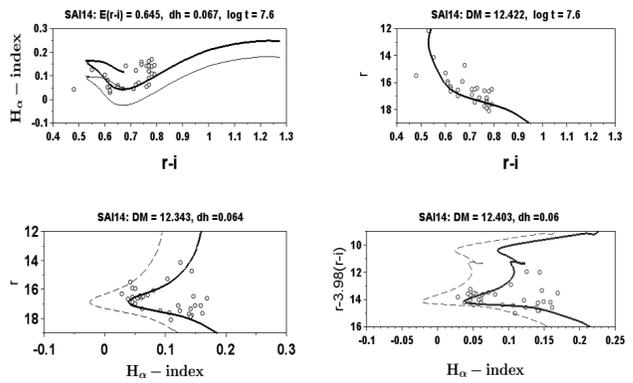}
\caption{Color-color diagram and the  ``color--magnitude diagram''  for proper-motion selected
cluster  members within 5~arcmin from the center of the SAI~14 cluster. The lines are the same as
in Fig.~\ref{fig:param_all}} \label{fig:param_mem}
\end{figure}

We then again determined the color excesses, distance moduli, and ages of 108 clusters using the above method
and applying it to likely cluster members. Fig.~\ref{fig:param_mem} shows the same diagrams
as in Fig.~\ref{fig:param_all}, but based on  $SAI~14$ cluster members exclusively.

\subsection{Extinction law}
\label{extinction}

When computing cluster distances one has to take into account interstellar extinction toward the
cluster. In most cases the so-called ``standard'' extinction law of Cardelli et al.~\citep{Cardelli} or
its modification proposed by O'Donnell~\citep{ODonnell} is
used to this end. According to this law extinction is a given function of wavelength and
parameter $R_{V}$ that is equal to the ratio of extinction in the $V$ band to the color
excess $E_{B-V}$ ($R_{V}$~=$A_{V}$/$E_{B-V}$), which is supposed to be constant throughout the
Galactic disk and equal to $R_{V}=3.1$. Actually interstellar extinction law (which is in this case
determined by the value of parameter $R_{V}$) varies strongly both with the direction in the
Galaxy and with distance~\citep{Fitzpatrick}.
In this paper to characterize extinction law we employ parameter $R_r$, which is equal to the
ratio of total extinction in IPHAS $r$ band to the color excess $E_{r-i}$ ($R_r$~=$A_{r}$/$E_{r-i}$), 
and it is this parameter that we determine for extinction toward each cluster and then use
when computing the cluster distances. To study extinction law, we compiled near-infrared
photometry in the filters $J, H, K_s$ (2MASS), $W1$ (WISE),  and $i, z, y$ (Pan-STARRS), which we
used to compute the corresponding apparent color indices. We did not use WISE $W2$-band data
(with a effective wavelength of $\lambda = 4.6$~$\mu$m) because of an extinction peak in  the
vicinity of $\lambda = 4.5$~$\mu$m~~\citep{Gontcharov}, which evidently does not fit the 
supposed monotonic power-law decrease of extinction with wavelength. Table~\ref{filters} 
lists the effective wavelengths of the photometric passbands employed.

\begin{table}[]
\caption{Filters} \label{filters}
\medskip
\begin{center}
\begin{tabular}{l c c}
  \hline
Survey & Filter & $\lambda_{eff}$, $\mu$~m \\
\hline
IPHAS & $i$ &  0.77 \\
\hline
& $i$ & 0.75 \\
   Pan-STARSS & $z$ & 0.87 \\
    & $y$ & 0.96 \\
    \hline
    & $J$ & 1.25 \\
    2MASS & $H$ & 1.65 \\
    & $K_s$ & 2.17 \\
    \hline
    WISE & $W1$ & 3.35 \\
    \hline
\end{tabular}
\end{center}
\label{filters}
\end{table}

To determine the intrinsic (true) color indices, we used theoretical isochrones~\citep{Chen1, Chen2, Tang}
and the ZAMS determined with the initial mass function of Kroupa~\citep{Kroupa}. We fitted 
the distribution of color excesses $E(r - \lambda_i)$ = $(r -
\lambda_i)_{vis}$ - $(r - \lambda_i)_0$ for stars in each cluster by a Gaussian function to 
determine the average $E(r - \lambda_i)$ values.

We then plotted for each cluster the dependence of the average  $E(r-\lambda_i)$ color excess
on effective wavelength $\lambda_{eff}$($i$). In the near infrared this dependence
can be described by a function of the following form~\citep{Drew}:

\begin{equation}
    \label{AbsLowEq}
        E(r-\lambda) = a + b\cdot{\lambda^{-\alpha}}
    \end{equation}

The color excess  $E(r-\lambda)$ in the limit $\lambda \rightarrow \infty$ tends to
$A_r$ -- the total extinction in the $r$-band filter, i.e., solving equation~(\ref{AbsLowEq}) 
yields $A_r = a$. Given the earlier determined color excess $E(r-i$)  (see
Section~\ref{parameters} above), we determine the parameter $R_r = \frac{A_r}{E(r-i)}$ of the
extinction law.

To determine the extinction law toward each cluster, we selected stars lying in the vicinity 
of the minimum in the  ($r$-$i$, $H_{\alpha}index$) diagram. In the case of 14 clusters out of 108
we found no or too selected stars with available $W1$-band magnitude, and performed all subsequent
computations for the remaining 94 open clusters. We refined the membership status of selected stars 
based on the proper motions from Gaia DR2 catalog, which became publicly available at this
stage of our study. We fitted the distribution of stars in the $(\mu_\alpha,\mu_\delta)$ plane
by a two-component Gaussian:

\begin{gather}
            \begin{split}
                f = \frac{1}{2\pi\sigma_\alpha\sigma_\delta\sqrt{1-k^2}}\cdot \exp\Big(-\frac{1}{2(1-k^2)}\Big[\frac{(\mu_\alpha^i-\mu_\alpha)^2}{\sigma_\alpha^2} - \\
                -k\frac{2(\mu_\alpha^i-\mu_\alpha)(\mu_\delta^i-\mu_\delta)}{\sigma_\alpha\sigma_\delta}+\frac{(\mu_\delta^i-\mu_\delta)^2}{\sigma_\delta^2}\Big]\Big),
            \end{split}
        \end{gather}

where $\mu_\alpha, \sigma_\alpha, \mu_\delta, \sigma_\delta$ -
are the means and standard deviations of the corresponding distributions and  $k$ is the correlation coefficient. 
The probability contours have the form of ellipses described by equation:

 \begin{equation}
        \frac{(\mu_\alpha^i-\mu_\alpha)^2}{\sigma_\alpha^2}-k\frac{2(\mu_\alpha^i-\mu_\alpha)(\mu_\delta^i-\mu_\delta)}{\sigma_\alpha\sigma_\delta}+\frac{(\mu_\delta^i-\mu_\delta)^2}{\sigma_\delta^2} = const
    \end{equation}

 The $const = 9$ ellipse defines the area of the scatter of proper motions of individual stars
within $3\sigma$ of the mean values, and it is these stars that we use to derive the extinction law,
which we approximated via non-linear least-squares fitting. After the first approximation we rejected
stars lying outside the  $3\sigma$ and repeated the procedure. We used the
inferred $\mu_\alpha, \sigma_\alpha, \mu_\delta, \sigma_\delta$  values to estimate the components of the average
proper motion of the cluster and their standard errors.

We then determined the parameters $R_r$ and $\alpha$ of the
interstellar extinction law from the data of identified cluster members and used the inferred $R_r$ 
value to compute the photometric distance to the cluster in pc and the corresponding photometric
parallax. We also computed the weighted mean trigonometric parallaxes of the clusters by averaging 
Gaia DR2 individual of cluster members.

\section{Results}
\label{Results}

We list the results in Table~\ref{ocldata}. Column~1 gives the name of the cluster;
columns 2 and 3, the photometric distance to the cluster and its standard error in pc, respectively.
Columns 4 and 5 give the photometric parallax of the cluster computed from the photometric distance 
and its standard error, respectively. Columns 6 and 7 give the astrometric (trigonometric) parallax 
of the cluster computed from Gaia DR2 data and its standard error, respectively. Columns 8 gives the
logarithm of the cluster age in Myr; columns 9 and 10
give the extinction-law parameter $R_r$ and its standard error, respectively; columns 11 and 12 
give the total extinction in the  $r$ band and its standard error, respectively; columns 13 and
14 give the exponent $\alpha$ from equation~(\ref{AbsLowEq}) and its standard error, respectively. 
Columns 15 and 16 give the average proper-motion component of the cluster along right ascension and its 
standard error in mas/yr, respectively. Columns 17 and 18 give average proper-motion component of the cluster 
along declination and its  standard error in mas/yr, respectively. The full version of Table~\ref{ocldata} 
is available in electronic form  at:
\newline
\verb|www.sai.msu.ru/groups/cluster/cl/iphas_ocl|

\begin{table*}[]
\scriptsize \caption{Parameters of open clusters} \label{Data}
\medskip
\begin{tabular}{l c c c c c c c c c c c c c c c c c}
  \hline
Cluster & D, & $\sigma_D$, & $\pi$,& $\sigma_{\pi}$, & $\pi_{gaia}$, & $\sigma_{\pi_{gaia}}$, & log(t), &$R_r$ & $\sigma_{R_{r}}$ & $A_r$ & $\sigma_{A_{r}}$ & $\alpha$ & $\sigma_{\alpha}$ &  $\mu_{\alpha}$,  & $\sigma_{\mu_{\alpha}}$,  &  $\mu_{\delta}$,  & $\sigma_{\mu_{\delta}}$, \\
 & pc & pc & mas & mas & mas & mas  & Myr &  &  &  &  &  &  & {mas/yr} & {mas/yr} & {mas/yr} & {mas/yr}  \\
\hline
sai13 & 3035 & 220 & 0.329 & 0.024 & 0.283 & 0.005  & 8.2 & 3.752 & 0.048 & 3.487 & 0.057 & 1.934 & 0.039 & -1.610 & 0.009 & -0.056 & 0.009  \\
sai14 & 2873 &  89 & 0.348 & 0.011 & 0.307 & 0.010  & 7.6 & 4.131 & 0.083 & 2.666 & 0.060 & 1.838 & 0.085 & -1.555 & 0.072 & -0.269 & 0.053  \\
\hline
\end{tabular}
\label{ocldata}
\end{table*}

\begin{figure}[]
\includegraphics[width=\linewidth]{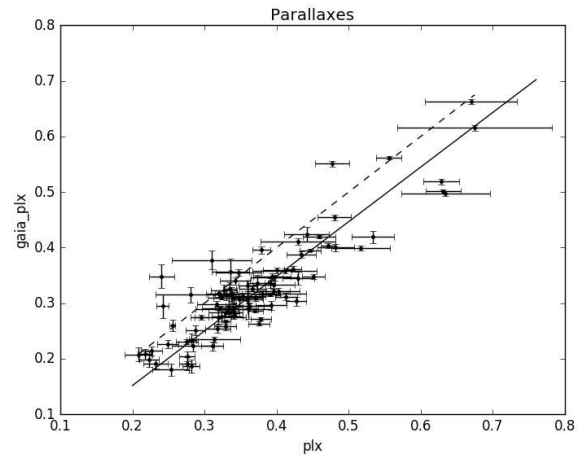}
\caption{Comparison of photometric and Gaia DR2 astrometric parallaxes. The dashed line
shows the equality diagonal and the solid line shows the inferred linear relation.}
\label{fig:comparep}
\end{figure}

Figure~\ref{fig:comparep} compares the photometric and Gaia DR2 astrometric parallaxes
of the clusters studied. As is evident from the figure, astrometric parallaxes are systematically 
smaller than photometric parallaxes. The dashed line shows the equality diagonal and the solid line shows the
linear relation found by minimizing the  function of the following form via the simplex method:

\begin{equation}
        \chi^2 = \sum\limits_{i=1} \frac{(\pi_{gaia,i} - a\cdot \pi_i - b)^2}{\sigma_{gaia,i}^2 + a^2 \cdot
        \sigma_i^2},
    \end{equation}

where the constants are equal to $ a_{opt} = 0.983 \pm 0.025$ and
$b_{opt} = -44.6 \pm 8.9$ microarcseconds. We  estimated the errors by constructing the section
of surface $\chi^2(a,b)$ by the plane $\chi^2 = \chi^2(a_{opt},b_{opt})+1$. Hence our photometric
distance scale (and the scale of photometric parallaxes) of clusters is consistent with the
distance scale implied by Gaia DR2 trigonometric parallaxes (the coefficient  $ a_{opt} = 0.983 \pm 0.025$ 
does not differ from unity within the quotet errors), and the systematic error of the scale of
Gaia DR2 trigonometric parallaxes is of about -44.6~$\pm$~8.9~microarcseconds (measured
trigonometric parallaxes are, on the average, underestimated by  this amount).

In their guidelines for using Gaia DR2 parallaxes Luri et al.~\citep{Luri} admit a zero-point offset of
30~microarcseconds. Zinn et al.~\citep{Zinn} estimate the Gaia DR2 parallax zero-point offset to be
$52.8 \pm 3.4$ and $50.2 \pm 3.5$~microarcseconds from the data for red-giant branch and red-clump stars,
respectively. Riess et al.~\citep{Riess} found the offset to be  $46 \pm 13$~microarcseconds from 
their analysis of bright Galactic Cepheids. Like in this study, Gaia DR2 parallaxes are always found to be
systematically underestimated.

\begin{figure}[]
\includegraphics[width=\linewidth]{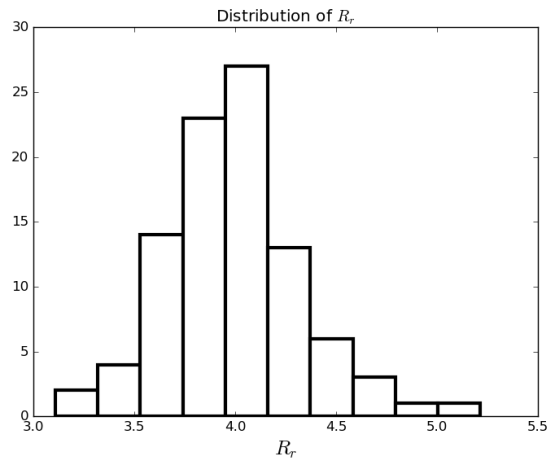}
\caption{Distribution of extinction-law parameter $R_r$ in the fields of 94 open clusters studied.} \label{fig:Rr}
\end{figure}

Figure~\ref{fig:Rr} shows the distribution of inferred values of parameter $R_r$,
which characterizes interstellar extinction law.
The mean value $<R_r> =  3.99 \pm 0.04$ differs significantly (albeit slightly)
from the value implied by the  ``standard'' law of Cardelli et al.~\citep{Cardelli} ($R_r = 3.88$),
agrees better with the extinction law of O'Donnell~\citep{ODonnell} ($R_r = 4.07$), both
with $R_{V}$~=3.1. Note the broad scatter of the distribution in Fig.~\ref{fig:Rr} with
a standard deviation of $\sigma_{R_r} = 0.35$. 

\begin{figure}[]
\includegraphics[width=\linewidth]{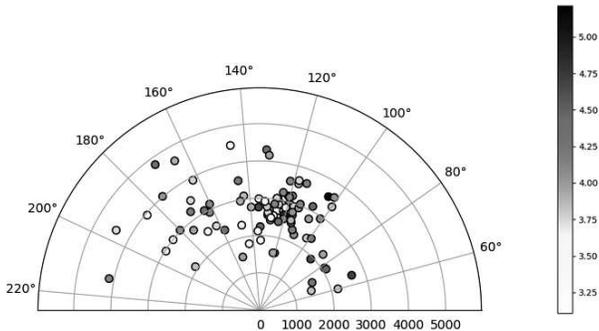}
\caption{Distribution of clusters projected onto the Galactic plane. The Sun is at
the coordinate origin, the radius gives the heliocentric distance in pc and the angle, the 
Galactic longitude. The color corresponds to the value of parameter $R_r$}. \label{fig:Rrpolar}
\end{figure}

Figure~\ref{fig:Rrpolar} shows the distribution of the clusters studied projected onto
the Galactic plane in polar coordinates. The color of the circles corresponds to the
inferred  $R_r$ value: the darker color, the higher  $R_r$. Extinction law can be seen to
vary conspicuously both with distance and direction.

\begin{figure}[]
\includegraphics[width=\linewidth]{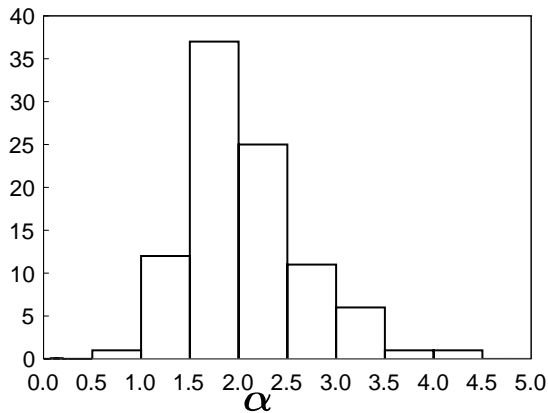}
\caption{Distribution of extinction-law parameter $\alpha$ in the fields of 94 open clusters studied.} \label{fig:alpha}
\end{figure}

In their review Matsunaga et al.~\citep{Matsunaga} point out that the studies performed
before 1995 yielded exponents $\alpha$ (equation~(\ref{AbsLowEq})) in the 1.6 -- 1.8 interval, 
whereas later estimates of $\alpha$ obtained by different authors span from 0.8 to 2.6~\citep{Moore}. 
The exponents  $\alpha$ found in this work span an even broader interval from 0.74 to 4.01 
(see Fig.~\ref{fig:alpha}) with a mean and standard deviation equal to
$<\alpha> = 2.05$ and $\sigma_\alpha = 0.58$, respectively. 

\begin{figure}[]
\includegraphics[width=\linewidth]{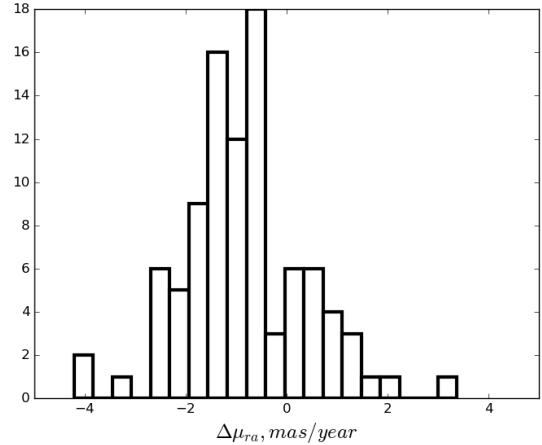}
\caption{Comparison of proper motions in right ascension based on positional data from seven catalogs 
(up to 13 different epochs per star spanning the time interval from 1949 through 2015) with the
proper motions based on Gaia DR2 data.} \label{fig:deltapmra}
\end{figure}

\begin{figure}[]
\includegraphics[width=\linewidth]{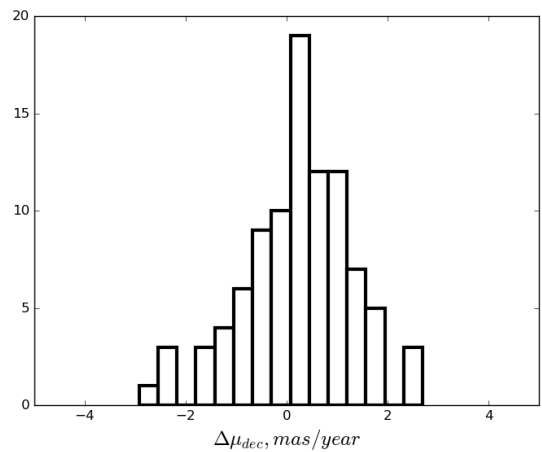}
\caption{Comparison of proper motions in declination based on positional data from seven catalogs 
(up to 13 different epochs per star spanning the time interval from 1949 through 2015) with the
proper motions based on Gaia DR2 data.}
\label{fig:deltapmde}
\end{figure}

Figure~\ref{fig:deltapmra}  shows the distribution of the
difference between the cluster proper-motion component in right ascension computed using
the Sanders method in Section~\ref{motions} and the cluster proper motions determined
from Gaia DR2 data in Section~\ref{extinction}. Figure~\ref{fig:deltapmde} shows the corresponding
difference for the proper-motion components in declination. The mean values and 
standard deviations are equal to $<\Delta{\mu_{\alpha}}> = -1.036 \pm 0.088 $~mas/yr, 
$\sigma{\Delta_{\mu_{\alpha}}} = 0.998 \pm 0.088 $~mas/yr and 
$<\Delta{\mu_{\delta}}> = 0.347 \pm 0.090 $~mas/yr, 
$\sigma{\Delta_{\mu_{\delta}}} = 0.910 \pm 0.091 $~mas/yr. Thus the mean differences
of proper-motion components prove to be comparable to the estimate of characteristic
systematic errors of the UCAC5 catalog, which we used to define our reference frame: 
according to UCAC5 authors, these systematic errors are within 0.7~mas/yr~\citep{UCAC5}. 
Table~\ref{ocldata} lists only the average cluster proper motions determined from
Gaia DR2 data.

\section{Conclusions}
\label{Conclusions} 
We determined the photometric distances,  $E_{r-i}$ color excesses (based on IPHAS survey data), ages,
the mean Gaia DR2 astrometric parallaxes and proper motions for almost one hundred Galactic open clusters. 
We estimated the photometric distances taking into account individual total-to-selective extinction
ratios $A_r$/$E_{r-i}$ for each cluster, which we determined using the color-difference method applied to 
IPHAS  survey $r$-, $i$-, and $H_{\alpha}$-band photometry, 2MASS $J$-, $H$, and $K_s$-band photometry, 
WISE $W1$-band photometry, and Pan-STARRS $i$-, $z$-, and $y$-band photometry. The inferred $R_r$ ratios vary significantly
from cluster to cluster, the mean and standard deviation are equal to <$R_r$>~=3.99 
and $\sigma R_r$~=0.34, respectively. 
We found our photometric distance scale to agree well with the scale of Gaia DR2 trigonometric parallaxes, which, 
on the average, are systematically underestimated by 45~$\pm$~9~microarcseconds in good agreement with
the results obtained by other authors based on other objects (classical Cepheids, quasars, red-giant and red-clump stars).

\section*{Acknowledgments}

This work was supported by the Russian Foundation for Basic Research (grant no.~18-02-00890).
This publication makes use of data products from the Two Micron
All Sky Survey (2MASS), which is a joint project of the University of Massachusetts
and the Infrared Processing and Analysis Center/California Institute of Technology,
funded by the National Aeronautics and Space Administration and the National Science Foundation,
and of the data products from the Wide-field Infrared Survey Explorer (WISE), which is a joint project
of the University of California, Los Angeles, and the Jet Propulsion Laboratory/California
Institute of Technology, and NEOWISE, which is a project of the Jet Propulsion Laboratory/California
Institute of Technology. WISE and NEOWISE are funded by the National Aeronautics and Space Administration.
This publication also makes use of the  data obtained as part of
the INT Photometric $H_{\alpha}$ Survey of the Northern Galactic
Plane (IPHAS, www.iphas.org) carried out at the Isaac
Newton Telescope (INT) and the data of Pan-STARRS 1 survey, as well as
the data from the European Space Agency (ESA)
mission {\it Gaia} (\url{https://www.cosmos.esa.int/gaia}), processed by
the {\it Gaia} Data Processing and Analysis Consortium (DPAC,
\url{https://www.cosmos.esa.int/web/gaia/dpac/consortium}). Funding
for the DPAC has been provided by national institutions, in particular
the institutions participating in the {\it Gaia} Multilateral Agreement.




\clearpage

\onecolumngrid \clearpage

\end{document}